\documentclass[usenatbib,usegraphicx]{mn2e}

\newcommand\Msun{\mbox{M$_{\odot}$}}%

\begin{document}

\title[Recycled pulsars with black hole companions]{Recycled pulsars with black hole companions: the high mass analogues of PSR~B2303$+$46}
 
\author[M. Sipior et al.]{M.~S.~Sipior,$^1$ S. Portegies Zwart,$^1$ G. Nelemans$^2$ \\
$^1$Astronomical Institute ``Anton Pannekoek'' and Section Computational
  Science, University of Amsterdam, \\ 
$\;$Kruislaan 403, 1098 SJ Amsterdam, Netherlands. \\
$^2$Institute of Astronomy, University of Cambridge, Madingley Road,
Cambridge CB30HA, UK}

\maketitle


\begin{abstract}

We investigate the possibility that mass transfer early in the evolution of
a massive binary can effect a reversal of the end states of the two
components, resulting in a neutron star which forms before a black hole. In
this sense, such systems would comprise the high-mass analogues of white
dwarf-neutron star systems such as PSR~B2303$+$46. One consequence of this
reversal is that a second episode of mass transfer from the black hole
progenitor star can recycle the nascent neutron star, extending the life of
the pulsar.

An estimate of the formation rate through this channel is first performed
via a simple analytical approach, and then refined using the results of the 
SeBa binary evolution package. The central role of kicks in determining the
survival rate of these binaries is clearly demonstrated. The final result is
expressed in terms of the number of field pulsars one can expect for every
single neutron star-black hole (ns,bh) binary. We also calculate this figure 
for black hole-neutron star (bh,ns) systems formed
through the usual channel. Assuming kicks drawn from the distribution of
Cordes \& Chernoff (1998), we find an expectation value of one (ns,bh) binary
per $4\times10^4$ pulsars, and one (bh,ns) system for every 1500 pulsars. This
helps to explain why neither system has been seen to date, though it
suggests that detection of a (bh,ns) binary is imminent.

\end{abstract}


\section{Introduction}

The ``recycling'' of pulsars through mass accretion from a binary
companion has been established in the literature for more than two decades
\citep{alpar:1982}.  In the case of neutron star-neutron star binaries,
the recycled pulsar is in fact the first neutron star to form, as the
mass accretion is driven by Roche-lobe overflow during the evolution
of the second, initially less massive star. There are several double
neutron star systems (ns,ns) currently known, which have all experienced
some degree of recycling \citep{arzoumanian:1999,burgay:2003}.

The second star need not end up as a neutron star, of
course. \citet{kerkwijk:1999} and \citet{davies:2002} showed that
the binary PSR B2303+46, which contains a slowly rotating pulsar, was
consistent with a white dwarf-neutron star (wd,ns) system where the
white dwarf evolved \emph{first}.  This reversal leaves the neutron
star with no source of matter to accrete, hence no recycling takes
place in this instance, and one is left with a white dwarf and an
unrecycled pulsar. The mechanism behind this reversal is detailed
in \citet{tutukov:1993}, \citet{zwart:1999} and \citet{tauris:2000}.
Essentially, as the primary evolves and fills its Roche lobe, it spills
so much mass onto the lighter companion that the evolutionary destinies of
the two stars are reversed. The initially massive primary has insufficient
mass to become a neutron star, ending up as a heavy white dwarf, whereas
the companion gains enough mass to evolve into a neutron star. We label
such a binary as a white dwarf-neutron star (wd,ns) system to distinguish
its heritage from the more common (ns,wd) binaries, in which the neutron
star forms first.  For the (wd,ns) evolution channel, the stars have to
be of similar mass, and these masses have to straddle or lie just below
the boundary between white dwarf and neutron star formation.

An analogous situation exists in higher mass systems, where both components
have a ZAMS mass below the threshold for black hole formation. Unlike the
(wd,ns) systems above, the pulsar in this scenario will experience 
recycling if the system undergoes Roche-lobe overflow when the
companion leaves the main sequence. As yet, no neutron stars have been
discovered with black hole companions. An
estimate of the rate at which such objects are formed, coupled with educated
guesses about the longevity of such a system, allows us to infer the number
of such systems that may exist in our Galaxy. A more useful comparison is the
ratio of the number of (ns,bh) systems to pulsars, as an expectation value for
the (ns,bh) detection rate per pulsar found. If this number is greater than the
number of pulsars currently known, the absence of (ns,bh) binaries is easily
understood. Conversely, if the rate implies that some (ns,bh) systems should
have been detected already, this estimate serves as a lower bound for
effects that select against the detection of these binaries.

\section{Analysis \label{analysis}}

To estimate the frequency of (ns,bh) systems in the Galaxy, we construct a
probability for a single binary system to be generated with the initial
parameters that are a necessary (but not sufficient) condition for
(ns,bh) formation. We assume that stars with masses between $8$ and
$25 \,\Msun$ become neutron stars, with stars above this limit becoming
black holes. This is consistent with recent work by \citet{fryer:2002}
and \citet{heger:2003} on the end state of massive stars. The binarity
fraction is taken to be 50\% (i.~e., 2 out of 3 stars are in a binary),
and a standard Salpeter IMF is assumed ($dN = m^{-2.35}\,dm$), drawn
from 8 to 100 M$_{\odot}$. The mass of the secondary is determined by
drawing the mass ratio $q$ from a flat distribution. The choice of a mass
ratio distribution is still quite contentious; a study of massive stars in
the Orion nebula \citep{preibisch:1999} showed a secondary IMF that is
consistent with that of the field star population, but a flat distribution
could not be ruled out either. Given the inherent observational biases in
studying systems with wide mass ratios \citep{hogeveen:1992}, a flat mass
ratio distribution is not an unreasonable assumption here. A distribution
skewed towards lower mass secondaries would, of course, result in a smaller
number of (ns,bh) systems.

To form a (ns,bh) system via this channel, a number of conditions must occur
in order, each with an associated probability. First, the system must
have a small enough initial separation for a first mass transfer episode.
Second, enough mass must be transferred from the primary to the secondary
to ensure that the secondary becomes a black hole, with a reduction in
its evolutionary timescale commensurate with the mass gained. The primary
must retain enough mass to become a neutron star. We assume that the main
sequence lifetime of the primary is not extended by mass loss (as it is
the envelope which is lost). Third, the system must survive the natal kick
that accompanies the collapse of the primary to a neutron star. Fourth,
the perturbed system must retain a sufficiently close orbit that a second
bout of mass transfer can occur, this time from the more massive secondary
onto the neutron star primary. Lastly, the system must survive any natal
kick arising from the collapse of the secondary into a black hole.

Setting aside for a moment the question of whether many (ns,bh) progenitor
systems survive the first natal kick, we can address in a basic, analytical way
the potential upper limit for the formation rate of these binaries.

The probability that an arbitrary binary will transfer mass as the
primary leaves the main sequence depends upon the ZAMS mass $M_0$ of the
primary, the initial mass ratio $q_0$, and the initial semi-major axis
$a_0$. Practically speaking, such systems have initial periods of less than
10 years \citep{pols:1991}, with semi-major axes on the order of a few
thousand solar radii. Initial orbital separations are typically taken to
be distributed evenly in $\log a$, out to roughly $10^6 R_{\odot}$
\citep{abt:1983}.
As the region of interest occupies three decades in $a$, out of
a possible six, we take the probability of initial mass transfer to be $1/2$.

We assume that the entire envelope of the primary is transferred during this
initial episode, though based on the degree to which the transfer is
non-conservative, not all of this matter will be accreted by the secondary.
With a mass loss fraction $f$, the final masses ($M_1, M_2$) of the 
components are related to their initial masses ($M^0_1, M^0_2$) by the
expressions
\begin{eqnarray}
M_1 & = & M_{core} \\
M_2 & = & M^0_2 + (M^0_1 - M_{core}) (1 - f)
\end{eqnarray}
where $M_{core}$ is the core mass of the primary, which we take to be $0.058
(M^0_1)^{1.57}\,\mbox{M}_{\odot}$ \citep{iben:1985}. 

The mass loss fraction $f$ is taken to be a
linear function of the initial mass ratio only \citep{pols:1991,zwart:1995},
ranging from $1$ at $q_0 = 0.2$ to $0$ when $q_0 = 0.6$. Values above or
below this range assume a value of unity or zero, respectively. An
alternative is to treat $f$ as a quadratic function of $q$; in the simplest
case, $f = 1 - q^2$. We will have more to say about the relevance of this
parameter later when discussing the results of a detailed binary evolution 
code in section \ref{discuss}.

The final probability to consider is whether the system will remain
sufficiently closely-bound to transfer mass from the black hole progenitor
back onto the nascent neutron star. To first order, we consider that any
system that survives the first natal kick will eventually go on to a second
bout of mass transfer. Given the extreme difference between the imparted
kick speed and typical orbital speeds for this scenario, those systems
receiving a prograde kick will almost invariably be disrupted, while systems
that remain intact may still be close enough for mass transfer after
tidal circularisation has completed.

The question of normalisation must also be considered. We choose to
normalise to the Galactic supernova rate of 0.01 yr$^{-1}$ (consistent
with the 0.008 yr$^{-1}$ reported by \citealt{cappellaro:1997}).
We assume all stars with a zero-age main sequence mass between 8 and 25 \Msun\
contribute to the supernova rate. The question of whether black holes
experience a supernova at formation is a complex and unresolved
one, with a critical dependence on the star's pre-collapse mass
\citep{fryer:2002}. For our purposes, we ignore the natal kick imparted
to a nascent black hole, and do not count these events towards the
supernova rate.

From the above considerations, we can establish the range of primary
and secondary masses for which the (ns,bh) scenario is physically
possible. Figure \ref{regions} shows the boundaries in mass space for
(ns,bh) candidates. The four boundary conditions are indicated by roman
numerals, and represent the following: (I) that the envelope mass $M_{e}$
of the primary is greater than the difference between the ZAMS mass $M_2$
of the secondary and 25 \Msun (the threshold for black hole formation),
so that $M_e = M_1 - 0.058 M_1^{1.57} \ge 25 - M_2$.  (II) The tautological
requirement that the initial primary mass is greater than that of the
secondary (i.~e., $0 < q \le 1$). (III) The ZAMS mass of the primary does
not exceed the threshold for black hole formation ($M_1 < 25 \Msun$). (IV)
If the initial mass transfer is very non-conservative, the envelope mass
$M_e$ will not be sufficient to make the secondary a black hole, so we
require that $M_2 \ge 25 - M_e (1 - f)$. Note that this last requirement
is essentially a more stringent version of condition (I), and only comes
into play for primaries above 20 \Msun.

A crude estimate of the (ns,bh) formation rate can be obtained by simply 
calculating the area bounded by the four aforementioned conditions, weighted
by the initial mass function. This gives an upper limit to the formation
rate, as it ignores the effects of natal kicks, and the likelihood of the
two mass transfer episodes. From our previous assumptions (Salpeter IMF,
flat mass ratio distribution, and a 50\% binarity), and normalising to the
Galactic SNR, we find a formation rate of $1.6 \times 10^{-4}\,$yr$^{-1}$.
If we consider a more severely nonconservative mass transfer, with $f = 1 -
q^2$, the formation rate shrinks by a factor of two, to
$7.5\times10^{-5}\,$yr$^{-1}$.

\section{Discussion}
\label{discuss}
Having determined the formation rate of recycled (ns,bh) binaries, we can
estimate the number of such systems per pulsar by comparing the product
of the respective formation rates and pulsar lifetimes. This is
dependent upon the extent to which the pulsar is recycled, a complicated
topic which we do not address at present. We take a lifetime $\tau_{l}
= 10^8\,$yr as a reasonable average, consistent with recycled pulsar
lifetimes inferred from neutron star binaries \citep{arzoumanian:1999}. For
unrecycled pulsars, we assume an average lifetime of $2\times10^7\,$yr. The
results are summarised in Table~\ref{numbers}.

The estimate of the (ns,bh) formation rate above is clearly too large; among
approximately 1500 known pulsars, many of these systems should have already
been seen. Most of the putative (ns,bh) systems should in fact have been
disrupted by symmetric and asymmetric kicks during the formation of the
neutron star. To better understand the extent to which kicks prevent these
systems from forming, we employ a more detailed numerical model of binary
evolution, including several choices regarding the distribution of natal
kick speeds.

We chose to use the SeBa package for our numerical model, as described
in \citet{zwart:1996}. However, several design choices require comment at this
point. Our prescription for handling helium star winds is that of
\citet{langer:1989}. The stellar evolution tracks that form the core of the
package are the analytic interpolations given in \citet{eggleton:1989}.

Probably the least well-constrained quantity in our model is the initial
black hole mass function; i.~e., the mapping of pre-collapse masses and the
mass of the resulting black hole. The approach used in SeBa is based on
that of \citet{fryer:2001}. There, the total binding energy of the
progenitor at collapse is compared to the energy released in the supernova,
multiplied by an efficiency factor which is taken to be roughly $1/2$. In
SeBa, we take the supernova energy as fixed at $10^{51}\,$ ergs, instead of
a function of the progenitor mass. The mass of the CO core is taken as the
minimum mass of the black hole, to which some fallback material may be
added. The binding energies for the helium and hydrogen envelopes are
calculated separately, and then each is compared to the supernova energy in
turn. If the hydrogen envelope has a mass $M_{H}$ with binding
energy $E_{H}$, a supernova providing energy $E_{SN}$ will
produce an amount of fallback hydrogen $F_{H}$ given by $F_{H} =
M_{H}\times(1-\frac{E_{SN}}{E_{H}})$ ($0 \le F_{H} \le 1$). Any
remaining supernova energy after the hydrogen layer is unbound is then
applied to the helium layer, so that the amount of fallback helium is
just $F_{He} = M_{He}\times(1 - \frac{E_{SN}-E_{H}}{E_{He}})$.  If any of the
hydrogen falls back on to the black hole, the entire helium envelope
will also be captured, by definition. The final black hole mass is then just 
the CO core mass plus the total fallback material.

To better investigate the role of kicks, we chose three separate
kick distributions.  The first allows a symmetric kick only, with a
zero-magnitude asymmetric kick. Next, we performed a series of runs
using the asymmetric kick distribution of \citet{paczynski:1990} and
\citet{hartman:1997}, with a dispersion speed of 300 km s$^{-1}$. Last
we use the model formulated by \citet{cordes:1998}, derived from
observations of the Galactic pulsar scale height. This last distribution
is a double-gaussian, drawing 86\% of the velocities from a gaussian
with a dispersion of $v_{disp} = 175$ km s$^{-1}$, and the remainder
from a gaussian with $v_{disp} = 700$ km s$^{-1}$.

The results of this simulation are summarised in Table \ref{numbers},
directly below the event rates derived from a direct integration as
described in the Analysis section above (sets A and B in the table). The
set C come from the output of the SeBa binary evolution code with no
asymmetric kicks applied at neutron star formation (symmetric kicks still
occur). D and E also come from the SeBa routine, but with natal kicks drawn,
respectively, from the Paczynski and Cordes-Chernoff distributions described
above. The (ns,bh) formation rate is found by scaling the output to the
Galactic supernova rate of 0.01 yr$^{-1}$. As well, it is assumed that
recycled pulsars will live an average of $10^8$ yr, as before, while slow
pulsars are taken to have a lifetime of $2\times10^7$ yr for this estimation.
We immediately see the enormous impact that natal kicks have on the
formation rates of such binaries. The considerable difference between the
analytical results (A and B), and the zero-kick numerical simulations
arises from ignoring the effects of wind mass loss in the former. Systems that
survive the first supernova may widen considerably due to wind mass loss
from the secondary as it leaves the main sequence, preventing a second mass
transfer episode \citep{zwart:1997}. Hence our initial overestimate of the
formation rate.

Given that the number of known pulsars is less than two thousand,
it is not at all surprising that no (ns,bh) system has been found to
date. While strongly dependant upon the chosen natal kick distribution,
it is clear that the total number of known pulsars must grow by a factor
of several before even one such binary is likely to be found.

\begin{figure}
\includegraphics[width=80mm]{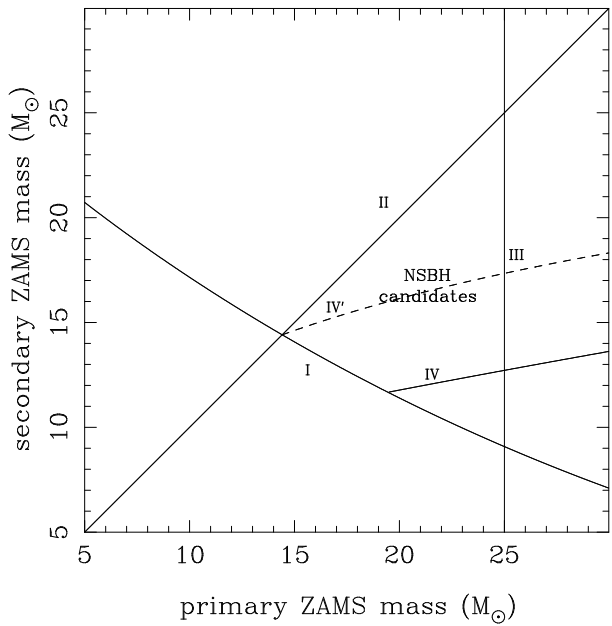}
\vspace{0.5cm}
\caption{Shows the range of stellar mass pairs that could conceivably lead
to a (ns,bh) end system. The four boundary conditions are delineated by Roman
numerals, and are derived from: (I) the requirement that a sufficient amount
of mass be transferred from primary to secondary to ensure that the latter
becomes a black hole, (II) the tautological requirement that the primary
ZAMS mass is greater than that of the secondary, (III) the
\emph{maximum} mass of the primary, above which the star must inevitably
become a black hole (25 \Msun), and (IV) the limit of non-conservative mass
transfer at low values of $q$. As mass transfer becomes less conservative, a
larger envelope mass is required to raise the secondary above the black hole
mass threshold. Condition (IV$^{\prime}$) results from an alternative
assumption of more severely nonconservative mass transfer.}
\label{regions}
\end{figure}

\begin{figure}
\includegraphics[width=80mm]{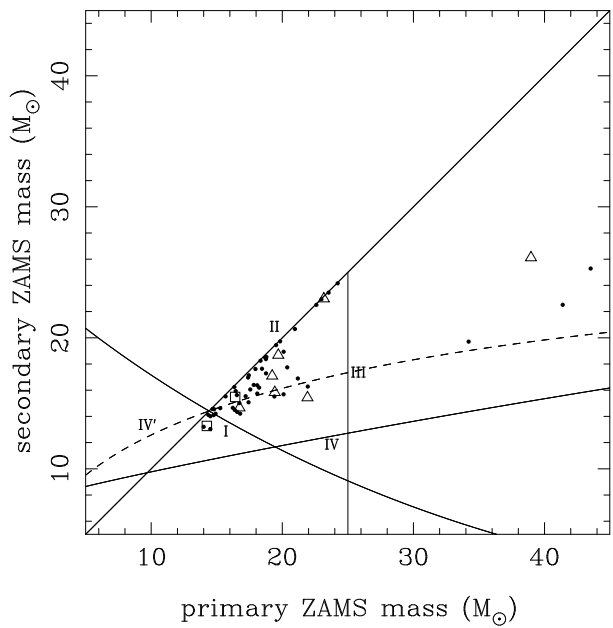}
\vspace{0.5cm}
\caption{Primary and secondary ZAMS masses for recycled (ns,bh) systems
generated by the SeBa stellar evolution code. Dots represent systems
formed when no asymmetric kicks apply; triangles and squares show
systems that survived a Paczynski- or Cordes-Chernoff-type natal kick,
respectively.  These systems are principally confined to the region
bounded by the four theoretical conditions. Outliers at low masses result
from a lower mass limit for black hole formation.  Several confirmed
instance of Case A mass transfer also resulted in the formation of
recycled (ns,bh) binaries (the high primary mass outliers), though our
putative scenario initially considered only Case B mass transfer for
this channel.}
\label{regions_seba}
\end{figure}

Figure \ref{regions_seba} shows the initial ZAMS masses of both components
of (ns,bh) progenitors. For each of the three choices of kick distribution,
$1.2\times10^6$ binaries were generated, with the remaining intact (ns,bh)
systems shown on the plot. We then overlayed the criteria from Figure
\ref{regions} to check how well our constraints were borne out. The bulk of
the (ns,bh) systems fall inside the theoretical bounds we established. One
exception is our assumption about the extent of nonconservative mass
transfer as a function of the mass ratio $q$. Clearly, our initial linear
approximation for the conservation factor $f$ was too lenient. The dashed
line in Figures \ref{regions} and \ref{regions_seba} represents the mass
conservation condition when $f = 1 - q^2$. This assumption better matches the
data, but the true curve lies between these extremes. Of special note are
the scattering of (ns,bh) systems with very high primary ZAMS masses, well to
the right of condition III in both figures. These rare systems do result in
(ns,bh) binaries, but are the result of Case A mass
transfer from a very massive primary to a secondary that is just below the
mass threshold for black hole formation. We did not consider Case A mass
transfer when formulating our initial scenario, but it is interesting to note
this alternate, albeit secondary formation channel. The fraction of (ns,bh)
systems arising from Case A mass transfer varies as a function of the
natal kick distribution. With no natal kicks, this channel comprises about
10\% of the total (ns,bh) population. This rises to roughly 30\% when
Cordes-Chernoff or Paczynski-type kicks are applied, since systems that
exhibit Case A mass transfer are closely-bound by definition, and more
likely to survive a natal kick.

It is also interesting to consider the ``normal'' formation channel of (bh,ns)
binaries; i.~e., where the black hole forms first. The final two columns of
Table \ref{numbers} show the formation rate of these systems, and an
estimate of the ratio of regular pulsars to (bh,ns) binaries. While such
systems form roughly two orders of magnitude more frequently than (ns,bh)
binaries, the shorter lifetime of the unrecycled pulsar means that (bh,ns) are
only about 20--30 times more common than the reversed channel. Looking at
the expectation value of pulsars to (bh,ns) systems, we note that the number
for set E is comparable to the current number of known pulsars, implying
that detection of such a system could occur in the very near future. 

\section{Conclusions \label{conclude}}

We investigated a channel for the formation of (ns,bh) binaries, where
a bout of mass transfer reverses the ordinary evolutionary order,
resulting in a neutron star which forms first. This neutron star can
then be spun up by a second episode of mass transfer from the black
hole progenitor star, resulting in a recycled pulsar orbiting a black
hole. The rapidly-spinning pulsar should live several times longer than
a typical pulsar. We then estimated the formation rate of such systems
via both a simple analytical calculation and then a more detailed binary
evolution code, and used assumptions about the relative lifetimes of
the recycled pulsars to calculate the expected ratio of field pulsars to
(ns,bh) systems. The result is that the current known pulsar population must
grow by factors of several before such a system is likely to be found,
despite containing a longer-lived recycled pulsar.

Using the same methods, we show that normal (bh,ns) binaries should be more
common than (ns,bh) systems by an order of magnitude, and that the expected
number of pulsars per (bh,ns) system is comparable to the current known pulsar
population. For both (ns,bh) and (bh,ns) binaries, the strongest single 
factor in determining the formation rate is the distribution from which natal
asymmetric kicks are selected.

{\emph{Acknowledgments:}}
This work was supported by the Royal Netherlands Academy of Sciences
(KNAW), the Dutch organization of Science (NWO) and by the Netherlands
Research School for Astronomy (NOVA). GN is supported by PPARC.


\begin{table*}
\begin{minipage}{150mm}
\caption{Summary of recycled (ns,bh) formation rates\label{numbers}}
\begin{tabular}{ccccccc}
\hline
Set & Kick Disp.$\Delta V$ & (ns,bh) yr$^{-1}$ & 
(ns,bh) Total & PSR/(ns,bh) & (bh,ns) yr$^{-1}$ & PSR/(bh,ns) \\
\hline
A & 0 & $1.6\times10^{-4}$ & 16000 & 13 & $6.5\times10^{-4}$ & 15 \\
B & 0 & $7.5\times10^{-5}$ & 7500 & 27 & $6.5\times10^{-4}$ & 15 \\
C & 0 & $9.3\times10^{-7}$ & 93 & 2200 & $1.4\times10^{-4}$ & 70 \\
D & 300 & $1.2\times10^{-7}$ & 12 & 17000 & $2.1\times10^{-5}$ & 470 \\
E & 175/700 & $5\times10^{-8}$ & 5 & 40000 & $6.5\times10^{-6}$ & 1500 \\
\hline
\end{tabular}

\medskip
Set A is the result of direct integration of the area
in figure \ref{regions} weighted by the IMF, with the mass loss via
nonconservative mass transfer given by a simple linear relation, where
B is the same calculation, with the mass loss factor varying as the
square of the mass ratio $q$; sets C, D and E arise from detailed binary
simulations using the SeBa package. Set C includes symmetric Blaauw kicks
but no asymmetric kicks. Set D allows for a random asymmetric kick drawn
from the distribution of \citet{paczynski:1990}, with a dispersion
of 300 km s$^{-1}$; set E draws its kick from the distribution of
\citet{cordes:1998}, a double-Gaussian with dispersions of 175 and 700
km s$^{-1}$, respectively.  The (ns,bh) formation rate gives the number
of recycled (ns,bh) systems per year, while the next column estimates the
total number of recycled (ns,bh) systems in our Galaxy, by multiplying the
formation rate by the lifetime of a typical recycled pulsar, taken here
to be $10^8$ yr. Column 5 shows the expected number of pulsars per single
(ns,bh) system. Here we assume that pulsars are formed at the supernova rate,
and live an average of $2\times10^7$ years without recycling. The final
two columns show the formation rate of (bh,ns) systems, and the number of
pulsars expected for every (bh,ns) system.
\end{minipage}
\end{table*}


\bibliographystyle{mn2e}
\bibliography{references}

\end{document}